\documentclass[aip,floats,superscriptaddress]{revtex4}

\usepackage{graphicx}
\usepackage{epstopdf}
\usepackage{amssymb}
\usepackage{amsmath,bm}
\usepackage{psfrag}
\usepackage{epsfig}
\usepackage{float}
\usepackage{bm}
\newcommand{\angstrom}{\mbox{\normalfont\AA}}

\begin{document}

\title{Vacancy-mediated suppression 
 of phase separation 
 in a model two-dimensional surface alloy by the difference of the atomic jump rates
}

\author{Mikhail Khenner\footnote{Corresponding
author. E-mail: mikhail.khenner@wku.edu.}}
\affiliation{Department of Mathematics, Western Kentucky University, Bowling Green, KY 42101, USA}
\affiliation{Applied Physics Institute, Western Kentucky University, Bowling Green, KY 42101, USA}

\begin{abstract}
\noindent

A vacancy-mediated collective diffusion model is used to compute a thermally-induced (spinodal) phase separation in a typical fcc bi-metallic 
surface alloy at low median concentration of vacancies, focusing on the effect of the ratio, $\Gamma=\Gamma_A/\Gamma_B$, of the jump rates to the 
vacant sites of the two types of atoms. 
The model is formulated for the diffusion of one atomic species and vacancies, employing the kinetic transport 
coefficients derived by Moleko \emph{et al.} (Phil. Mag. A 59, 141 (1989)). It is demonstrated that for A$_{0.7}$B$_{0.3}$ alloy, $\Gamma \sim 50$ results 
in suppression of phase separation, whereas very small $\Gamma$ values result in mild phase separation (i.e., the suppression is incomplete).

\end{abstract}

\date{\today}
\maketitle


\section{Introduction}
\label{Intro}

Bi-metallic surface alloys are important materials for catalysis \cite{S}, since they can exhibit an improved catalytic activity through 
tailoring the concentration and/or the arrangement of two metallic components. Design of surface catalysts can benefit from surface phase-separated states
for enhanced area-selective catalytic activity \cite{FLKHYTHK}. Moreover, certain metal/semi-metal surface alloys are topological insulators, 
where ``the quantum confinement effect can suppress the bulk conduction channels and enhance the transport of the topological surface states" \cite{XLDGCGGMWPJXC}.
This property is valuable for design of future electronic and spintronic devices  \cite{SSFS,DUN}. 
If present, surface phase separation significantly affects surface electrical conductivity and therefore, it is bound to interfere with device performance.

For bimetallic surfaces, it is often difficult to determine, based on purely thermodynamic (energy-based) arguments, whether phase separation occurs 
\cite{RSN,CRSJSNB,Wandelt}, in part because a typical surface at a temperature significantly below its melting temperature is metastable, i.e. 
a non-equilibrium stationary structure may be kinetically stabilized by the presence of high energy barriers hindering the transition 
to an equilibrium state \cite{BEFBK}. On the other hand, in metal surfaces, structural rearrangements typically occur through vacancy-mediated 
diffusion \cite{GSASF,GSBH,AAFS,GSSF,Wandelt}. Therefore, understanding kinetic factors influencing vacancy-mediated diffusive phase separation 
in binary surface alloys is important, but we are not aware of any experiment or model/computational study that addresses this issue. 

The ratio of the jump rates to the vacant sites of the two types of atoms in a binary alloy, $\Gamma=\Gamma_A/\Gamma_B$,
is the most important kinetic factor governing vacancy-mediated diffusion.  
 In this communication, we fill the above-stated gap in knowledge by developing a ``collective", or ``chemical" (i.e., a continuum) vacancy-mediated 
diffusion model and the corresponding computation, aiming to answer a single question: what is the effect of $\Gamma$ on phase separation in a 
bi-metallic surface alloy ? 
 
The physical assumptions that underlie our model are customary \cite{VB}: 
(i) the jump rates are constant, i.e. they do not depend on the occupation of the sites surrounding the vacant sites; 
 (ii) the interactions between nearest neighbors atoms are pairwise, and (iii) the vacancies are conserved, which means that an atom from a surface monolayer cannot hop on top of 
a monolayer, thus creating a vacant site inside the monolayer, or an adatom executing surface diffusion cannot jump into a vacant site inside the monolayer. 
The consequence of this is that A and B atoms are conserved in the monolayer. Validity of the above assumptions has been confirmed by DFT calculations based on 
transition state theory \cite{MG}. 

\section{The Model}
\label{Model}

We consider a thin film of a uniform composition A, whereby the impurity B atoms are mixed into the topmost surface layer, thus creating
a two-dimensional binary surface alloy. Vacancy-mediated diffusion of $A$ and $B$ is assumed to take place in the atomically thin surface layer.

Since the model equations are somewhat cumbersome, we will state them immediately in the dimensionless form.
 Let $X_i(x,y,t)$ be the concentration, or the composition fraction, of the $i$-th species, defined as the number of atoms per unit area. Since each lattice site is occupied by an atom 
or a vacancy, the lattice site fraction is related to a concentration by $X_i=\nu C_i$, where $\nu$ is the lattice site density, having the same 
units as $X_i$, and $C_i(x,y,t)$ is the dimensionless concentration. In a perfect lattice, the number of lattice sites is conserved: $X_A+X_B+X_V=\nu$.
Dividing this relation by $\nu$ gives the constraint on the dimensionless concentrations:
\begin{equation}
C_A(x,y,t)+C_B(x,y,t)+C_V(x,y,t)=1. 
\label{sumC}
\end{equation}

 Next, we choose the lattice spacing $a$ as the length scale
and $\bar t=k T\nu/\left(\gamma_B \lambda \Gamma_B \right)$ as the time scale. Here $k T$ is Boltzmann's factor, $\gamma_B$ $(\gamma_A)$ the surface energy 
of atomically thin surface layer that is composed of B (A) atoms, $\lambda$ the 
dimensionless geometric factor for fcc lattice, and $\Gamma_B$ the surface jump rate of B atoms, i.e. the jump rate of B-V exchanges.
The following dimensionless parameters are introduced: $G=\gamma_A/\gamma_B$, the ratio of the surface energies of the atomic components; 
$N=k T\nu /\gamma_B$: the alloy entropy; $\xi=\delta /a \gamma_B$: the Cahn-Hilliard gradient energy coefficient 
(where $\delta$ is the dimensional gradient energy coefficient);  $\Gamma=\Gamma_A/\Gamma_B$: 
the ratio of the surface jump rates;
$H_{i,j}=\epsilon_{i,j}/\gamma_B,\ i,j=A,B$: the ratio of the bond strength between atoms of the same or a different kind to the B-surface energy;
and $Q>-1$: the quench parameter, defined by $T=T_c/(1+Q)$, where $T_c$ is the critical (spinodal) temperature. 
Positive $Q$ values correspond to $T<T_c$, which is the interval of the temperatures where phase separation is expected.

Our approach is based on the general continuum framework for vacancy-mediated 
 diffusion in binary alloys, as described by Allnatt \& Lidiard \cite{AllnattLidiard}. 
Note that although there are three concentration variables $C_A,\ C_B$ and $C_V$, due to the constraint (\ref{sumC}), only two of them are independent.
Therefore in the collective diffusion framework only the transport of one atomic species (B in our model) and vacancies is computed; 
the concentration field of the second atomic species (A in our model) is determined from Eq. (\ref{sumC}). 

The diffusion equations for $C_B$ and $C_V$ will follow from mass conservation:
\begin{equation}
\frac{\partial C_B}{\partial t} =   - \bm{\nabla} \cdot \bm{J}_B,\quad 
\frac{\partial C_V}{\partial t} =   - \bm{\nabla} \cdot \bm{J}_V, \label{C-eq}
\end{equation}
where 
$\bm{J}_B,\ \bm{J}_V$ are the diffusion fluxes, 
and $\bm{\nabla}=(\partial_x, \partial_y)$.  

Let $L_{AA},\ L_{AB},\ L_{BA}$ and $L_{BB}$ be the kinetic transport coefficients of the alloy components, $\mu_A,\ \mu_B$ and $\mu_V$ the chemical potentials of the 
alloy components and vacancies. According to Allnatt \& Lidiard \cite{AllnattLidiard} (p. 166, 174, 175)
\begin{eqnarray}
\bm{j_B} &=& - \left[L_{BV} \bm{\nabla}\left(\mu_V-\mu_A\right) + L_{BB} \bm{\nabla}\left(\mu_B-\mu_A\right)\right],\label{j-eq1} \\
\bm{j_V} &=& - \left[L_{VV} \bm{\nabla}\left(\mu_V-\mu_A\right) + L_{VB} \bm{\nabla}\left(\mu_B-\mu_A\right)\right],\label{j-eq2}
\end{eqnarray}
where 
\begin{equation}
L_{BV} = -\left(L_{BA} + L_{BB}\right),\ L_{VB} = -\left(L_{AB} + L_{BB}\right),\ L_{VV}=L_{AA} + L_{AB} + L_{BA} + L_{BB}
\end{equation}
are the effective kinetic transport coefficients after the diffusion of the $A$ species is eliminated.

The kinetic transport coefficients read:
\begin{equation}
L_{AA}=\Gamma C_A \left(1-\frac{2\Gamma C_B}{\psi}\right),\quad 
L_{AB}=\frac{2}{\psi}\Gamma C_A C_B,\quad L_{BA}=L_{AB},\quad
L_{BB}=C_B \left(1-\frac{2 C_A}{\psi}\right), \label{Ls}
\end{equation}
where $\psi$ is given by
\begin{equation}
\psi = \frac{1}{2}\left(M_0+2\right)\left(\Gamma C_A+C_B\right)-\Gamma-1+2\left(C_A+\Gamma C_B\right)+
\sqrt{\left[\frac{1}{2}\left(M_0+2\right)\left(\Gamma C_A+C_B\right)-\Gamma-1\right]^2+2M_0\Gamma}. \label{psi}
\end{equation}
In Eq. (\ref{psi}), 
$M_0=2f_0/\left(1-f_0\right)$, where $f_0$ is the tracer correlation factor for fcc lattice. 
Eq. (\ref{Ls}) stipulates that $L_{AA} \gg L_{BB}$ at $\Gamma \gg 1$, or $\Gamma_A \gg \Gamma_B$ 
(thus in this case A is the fast diffusing species); conversely, at $\Gamma \ll 1$, B is the fast diffusing species. At $\Gamma=1$ (the ``neutral" value)
and $C_A=C_B=0.5$, $L_{AA}=L_{BB}$. 
Eqs. (\ref{Ls}) and (\ref{psi}) are the dimensionless versions of 
the kinetic transport coefficients 
derived by Moleko et al. \cite{MAA}, who extended earlier results by Manning \cite{Manning1}. 
Allnatt and Lidiard \cite{AllnattLidiard} note that ``... although (the coefficients) were derived ...
with a specific model (the ``random alloy model") in mind the derivation given here is not obviously limited to that model. Furthermore, the argument is in no way 
limited to small vacancy fractions ... ." 
Additionally, as noted in Ref. \cite{Mehrer}, computations by Murch and co-workers \cite{BM1,BM2} also point to validity of 
 these relations for 
ordered alloys. (Further discussion of the relations (\ref{Ls}) and (\ref{psi}) can be found in Ref. \cite{VYCT}.)

The chemical potentials in Eq. (\ref{j-eq2}) follow from alloy thermodynamics \cite{ZVD,Iwai}:
\begin{eqnarray}
\mu_A &=& C_B\left(\frac{\partial \gamma}{\partial C_A}-\frac{\partial \gamma}{\partial C_B}\right)- \xi \bm{\nabla}^2 C_A - \left(1-Q^{-1}C_V\right) C_A,\label{muA} \\
\mu_B &=& -C_A\left(\frac{\partial \gamma}{\partial C_A}-\frac{\partial \gamma}{\partial C_B}\right)- \xi \bm{\nabla}^2 C_B + \left(1-Q^{-1}C_V\right) C_B,\label{muB} \\
\mu_V &=& Q\left(C_A^2+C_B^2\right)-\frac{\partial \gamma}{\partial C_V}=Q\left(C_A^2+C_B^2\right)+N \ln \frac{1-C_V}{C_V},\label{muV}
\end{eqnarray}
where the free energy
\begin{equation}
\gamma = G C_A + C_B + \frac{1}{2}\left(H_{AA} C_A^2 + H_{BB} C_B^2 + 2 H_{AB} C_A C_B\right) + 
N\left[C_A\ln C_A+C_B\ln C_B+C_V\ln C_V+\left(1-C_V\right)\ln \left(1-C_V\right) \right].
\label{gamma}
\end{equation}
In Eq. (\ref{gamma}) the first two terms constitute the weighted surface energy of the bi-component monolayer \cite{RSN}, whereas the terms in the 
parenthesis and bracket are the enthalpic \cite{VB} and entropic \cite{Iwai} contributions, respectively.

After $C_A$ is eliminated from Eqs. (\ref{C-eq})-(\ref{muV}) using Eq. (\ref{sumC}), the mathematical model is finalized. 
$C_A$ is determined as, 
$C_A=1-C_B-C_V$ \emph{after} $C_B$ and $C_V$ were computed using Eqs. (\ref{C-eq}).

Although the model may be applied to a rather general binary surface alloy, the material parameters that we use 
in Sections \ref{LSA} and \ref{CompositionEvolve} (see Table \ref{T1}) are closely matching fcc AgPt(111) surface alloy, where Ag is the deposited impurity species (B atoms in the model). 
Ref. \cite{RSBK} describes this surface alloy as
``... a solid solution with a positive enthalpy of mixing, which leads to phase separation". 
These authors found that deposition of Ag on Pt(111) and short annealing at a temperature slightly above 620K
results in a compact Ag clusters of the average size 10$\angstrom$ (about three lattice spacings across) embedded in the topmost Pt layer. 
Such partial de-mixing within the surface layer is 
consistent with the phase diagram for a bulk AgPt alloy, which features a wide miscibility gap up to $T\sim1400$K \cite{SCP}. 
The average 10$\angstrom$ size of two-dimensional Ag clusters is attributed to the balance of a lattice mismatch strain and the line (boundary) tension 
in the late stages of phase separation of the intermixed surface layer (the lattice mismatch of Ag and Pt is a moderate 4\% value).  
In the initial and intermediate stages of phase separation the effect of strain cannot be discerned in the experiment, thus this effect is presumably minor. 
For that reason, and since the model's 
continuum nature allows to compute only the initial and intermediate stages of phase separation (see Sec. \ref{CompositionEvolve}; this is sufficient to 
quantify the $\Gamma$-effect), we did not incorporate the strain into our model.

\begin{table}[!ht]
\centering
{\scriptsize 
\begin{tabular}
{|c|c|}

\hline
				 
			\rule[-2mm]{0mm}{6mm} \textbf{Physical parameter}	 & \textbf{Typical value}   \\
			\hline
                        \hline
			
                 \rule[-2mm]{0mm}{6mm} $a$ & $2.89\times 10^{-8}$ cm \cite{RSBK}   \\
			\hline
				\rule[-2mm]{0mm}{6mm} $\nu=a^{-2}$ & $9.18\times 10^{14}$ cm$^{-2}$   \\
			\hline
            \rule[-2mm]{0mm}{6mm} $\lambda$ & $1/6$ \cite{Manning1} \\
                        \hline
		    \rule[-2mm]{0mm}{6mm} $f_0$ & $0.7815$  \cite{Manning1} \\
                        \hline
			\rule[-2mm]{0mm}{6mm} $T$ & 400 K \cite{AAFS} \\
			\hline
			    \rule[-2mm]{0mm}{6mm} $\gamma_A$ & $2.3\times 10^3$  erg$/$cm$^2$ \cite{VRSK}   \\ 
				\hline
				\rule[-2mm]{0mm}{6mm} $\gamma_B$ & $1.17\times 10^3$  erg$/$cm$^2$ \cite{VRSK}   \\
				\hline
				\rule[-2mm]{0mm}{6mm} $\epsilon_{AB}=0.1\gamma_B$ & $117$  erg$/$cm$^2$   \\ 
			\hline
			\rule[-2mm]{0mm}{6mm} $\epsilon_{AA},\ \epsilon_{BB}=-0.1\gamma_B$ & $-117$  erg$/$cm$^2$   \\ 
			\hline
			    \rule[-2mm]{0mm}{6mm} $\delta$ & $1.2\times 10^{-5}$ erg$/$cm  \cite{Hoyt}  \\
			\hline

\end{tabular}}
\caption[\quad Physical parameters]{Physical  parameters. 
}
\label{T1}
\end{table}
\subsection{Notes on implementation}
\label{Implement}

The model as described above is still too complex for analysis and computation, due to the fact that the kinetic transport coefficients 
(and also the effective transport coefficients) are the highly nonlinear and cumbersome functions
of the space-time dependent concentrations $C_B$ and $C_V$.  Therefore, we approximate the kinetic transport coefficients by evaluating 
$\psi$ (Eq. (\ref{psi})) at $C_B=C_{B0},\ C_V=C_{V0}$, where $C_{B0}$ and $C_{V0}$ are the median concentrations in the as-prepared surface alloy, that is, 
the initial values for the simulation. 
This approximation has a very mild effect on the kinetic transport coefficients in the entire unit interval of the $C_B$ and $C_V$ values \cite{MyPhysRevMater} and moreover, 
in order to verify that results are not affected, a few simulations were done where the computation is periodically suspended, $\psi$ is re-computed 
using the instantaneous median values of $C_{B}$ and $C_{V}$, and the computation is resumed.     

Notice that the term $N \ln \frac{1-C_V}{C_V}$ in Eq. (\ref{muV}) diverges at $C_V=0,1$. A similar term, $N \ln \frac{1-C_B}{C_B}$, 
emerges in Eqs. (\ref{muA}) and (\ref{muB}) when in $\partial \gamma/ \partial C_A$, $\ln \left(1-C_B-C_V\right)$ is approximated 
as $\ln \left(1-C_B\right)$. This is valid as long as $C_V$ is small, the requirement that holds very well in the computation 
(see Sec. \ref{CompositionEvolve}; this is the only assumption of $C_V$ being small in the entire model formulation.) 
However, regularization of these logarithmic potentials is not needed, since 
the diffusion fluxes, Eqs. (\ref{j-eq1}) and (\ref{j-eq2}), are proportional to the gradients of the potentials. The gradients also are divergent at 0 and 1.
Thus to facilitate computation, the divergent
 gradients 
 are tapered by regularizing them as follows:
\begin{equation}
\bm{\nabla} \ln \frac{1-C_i}{C_i} =  \frac{\bm{\nabla} C_i}{C_i\left(C_i-1\right)} \approx
\frac{\bm{\nabla} C_i}{\left(C_i+\beta\right)\left(C_i-1-\beta\right)},\ i = B, V
\label{smoothlogs}
\end{equation}    
where $0<\beta\ll 1$. The chosen value for the computation is $\beta=0.01$. 

Note that for the example system, e.g. Ag/Pt(111) surface alloy, the assumption of small $C_V$ can be validated by computing the equilibrium vacancy 
concentration using the formula $n_v^{eq}=n \exp{(-q_v/kT)}$, where $n_v^{eq}$ is the number of vacancies, $n$ the number of surface sites, and $q_v$ a vacancy formation energy.
For Pt(111) surface $q_v=0.62$ eV \cite{UL}, for Ag(111) surface $q_v=0.67$ eV \cite{PMS}. Thus for AgPt(111) surface one can assume the average value $q_v=0.645$ eV.
Using this value, for the range of temperatures 300K-700K one obtains $1.46\times 10^{-11}\le n_v^{eq}\le 2.271\times 10^{-5}$ at $n=1$, which means that only 
one surface atom out of $6.85\times 10^{10}$ (at $n_v^{eq}=1.46\times 10^{-11}$), or out of $4.4\times 10^4$ (at $n_v^{eq}=2.271\times 10^{-5}$) 
is missing. Similar estimates of $n_v^{eq}$ were obtained by the embedded atom model \cite{GSASF}.

The computations in Sec. \ref{CompositionEvolve} use  
larger mean concentration $C_{V0}=0.01$, since at 
significantly smaller values a computation based on a continuum model breaks down due to divergence of $\bm{\nabla} C_V/C_V\left(C_V-1\right)$ 
at $C_V\rightarrow 0$, which even the mathematical regularization described above does not reliably eliminate. 
It follows from the linear stability analysis in the next section and from computations that the $\Gamma$-effect is not 
sensitive to the magnitude of $C_{V0}$. Essentially, only the computation time is affected when $C_{V0}$ is varied.  Moreover, 
according to Ref. \cite{VB} the statistical properties of vacancy-mediated diffusion (such as the phase boundary for the order-disorder transition and tracer diffusion coefficients) 
in the limit $C_V\rightarrow 0$ are accurately reproduced if the vacancy concentration 
is on the order of a few percents. That paper adopts $C_{V0}=0.04$ for kinetic Monte-Carlo computations.

Global conservation of $C_B$ and $C_V$ was checked during computation. The largest loss that occurred is 0.4\%. 

\subsection{Linear stability analysis}
\label{LSA}

\begin{figure}[H]
\vspace{-0.2cm}
\centering
\includegraphics[width=5.0in]{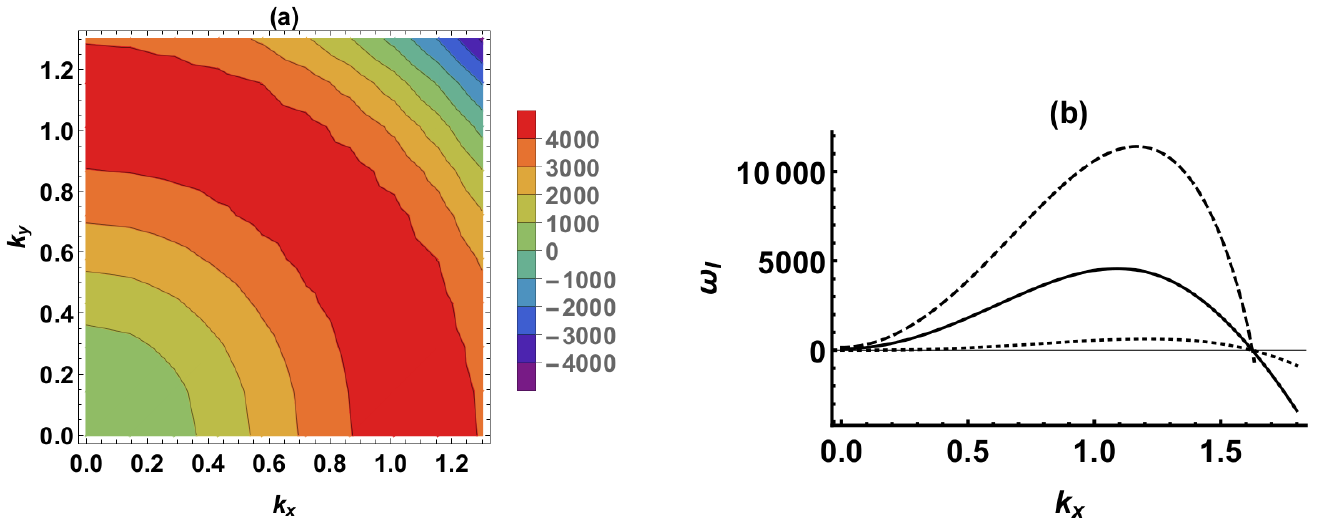}
\vspace{-0.15cm}
\caption{(a) The contour plot of the perturbation growth rate, $\omega_l\left(k_x,k_y\right)$, at $Q=0.005$ and $\Gamma=1$. (b) The cross-section plots, 
$\omega_l\left(k_x,1\right)$ at $Q=0.005$ and $\Gamma=1$ (solid line), $\Gamma=2$ (dashed line), $\Gamma=0.1$ (dotted line). 
}
\label{Fig1}
\end{figure}

From the linearized model equations in Sec. \ref{Model}, one obtains a quadratic equation for the growth rate $\omega(k_x, k_y)$ of the infinitesimal 
perturbations of a constant base state $C_B=C_{B0},\ C_V=C_{V0}$. Here $k_x,\ k_y$ are the perturbation wavenumbers.
 When both roots of this quadratic are negative for all $k_x,\ k_y$, 
the phase separation of the alloy does not occur.
In this case all small perturbations of the initial condition $C_B=C_{B0},\ C_V=C_{V0}$ decay.
Phase separation emerges due to growth of the perturbations when one or both roots are positive in a certain domain of the perturbation wavenumbers.
The largest positive root $\omega_l$ determines the domain of the unstable wavenumbers and the most dangerous wavenumber pair that corresponds to the 
maximum of the perturbation growth rate.\footnote{The expression for $\omega_l$ is too cumbersome to be presented in a paper or in supplemental materials. 
It was obtained with the help of a computer algebra system Mathematica. Mathematica notebook is available on request from the author.} As shown in Fig. \ref{Fig1} the instability has a long-wavelength character and the $\omega_l(k_x,k_y)$ 
surface is symmetric over $k_x-k_y$ plane. For the computations in Sec. \ref{CompositionEvolve} the size of the 
square computational domain is always chosen equal to $5\lambda_{max}$, where $\lambda_{max}=2\pi/k_x^{max}=2\pi/k_y^{max}$ and $k_x^{max}=k_y^{max}$ 
is the pair of the most dangerous wavenumbers. 

It can be noticed in Fig. \ref{Fig1}(b) that $\omega_l^{max}$ is more sensitive to increasing $\Gamma$ values, than to decreasing them. 
$\lambda_{max}$ is not very sensitive to changing $\Gamma$, it slightly increases when $\Gamma$ either decreases or increases from unity 
($\lambda_{max}=8.58,\ 7.74,\ 8.28$ at
$\Gamma=0.1,\ 1,\ 2$, respectively).
\section{Computational results}
\label{CompositionEvolve}

\begin{figure}[H]
\vspace{-0.2cm}
\centering
\includegraphics[width=2.5in]{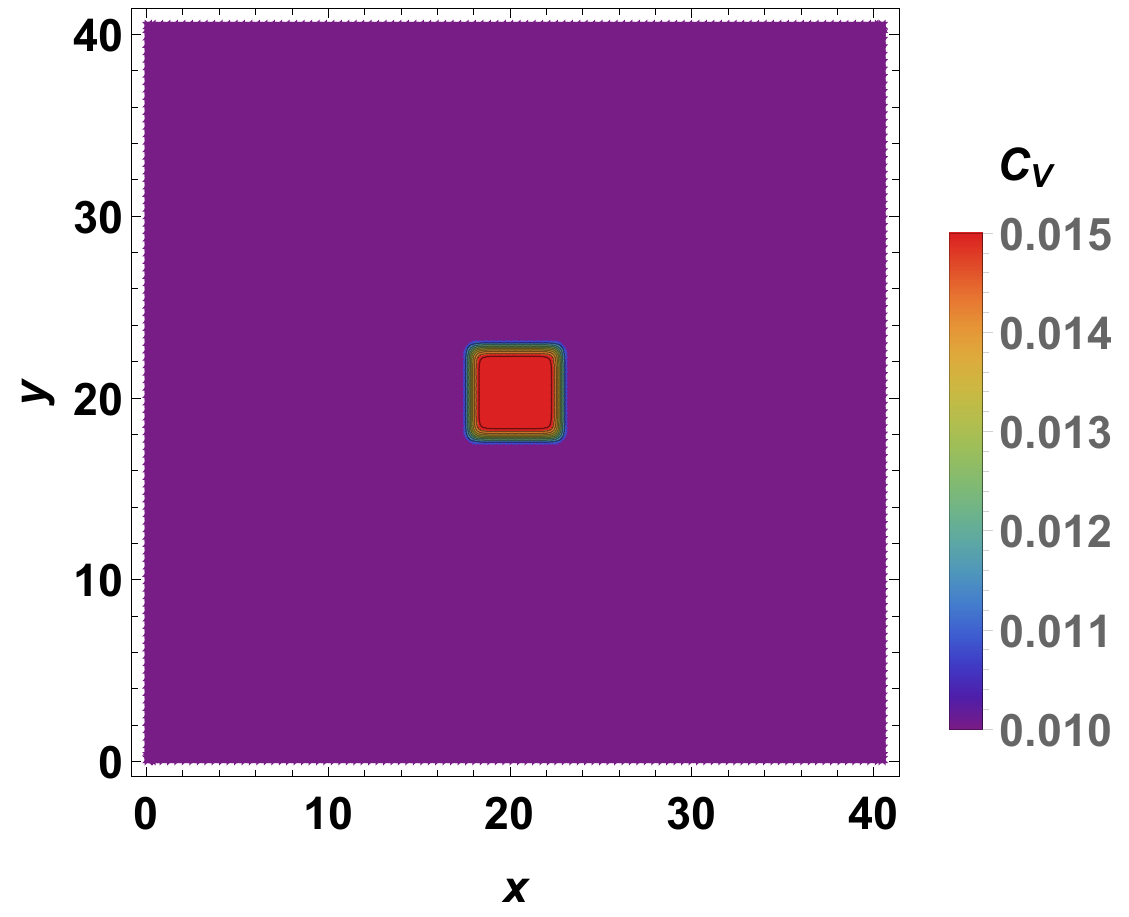}
\vspace{-0.15cm}
\caption{The vacancy concentration field at $t=0$.
}
\label{Fig2}
\end{figure}

Computations of Eqs. (\ref{C-eq}) are performed on the square, bi-periodic domain of the size $5\lambda_{max}$.  
The initial condition is chosen as the constant concentration field of B atoms and a square vacancy island on the background 
of a very low constant concentration field of vacancies, that is, 
$C_B(x,y,0)=C_{B0}=0.3,\ C_V(x,y,0)=C_{V0}+\eta(x,y)=0.01+\eta(x,y)$, where $\eta(x,y)$ is the "square box" function of the height 0.005 with a local support at the center 
of the computational domain. See Fig. \ref{Fig2}. Such initial condition may be engineered by a low-energy focused ion beam irradiation of a surface, 
or by using other local tools to remove atoms from a surface, such as a tip of an atomic force microscope or a scanning tunneling microscope \cite{WTW}. When the size of the computational domain changes due to $\lambda_{max}$ change in response to
a changes in the values of the material parameters $\Gamma$ and $Q$, the local support of the island is changed to preserve the relative size of the island. 
(Notice that the initial data above and the fact that $C_{A0}=1-C_B(x,y,0)-C_V(x,y,0)$ yield $0.685\le C_{A0}\le 0.69$.)

As the computation of the phase separation progresses, the concentration of vacancies may vanish in some sub-region. In our setup, this typically happens 
fast due to the initial vacancy concentration being small already. 
Such event corresponds to the local composition attaining a value pair $C_{Blocal}$ and $C_{Alocal}=1-C_{Blocal}$. In this case the computation is 
aborted and declared completed, and the concentrations are deemed final. 

\begin{figure}[H]
\vspace{-0.2cm}
\centering
\includegraphics[width=6.5in]{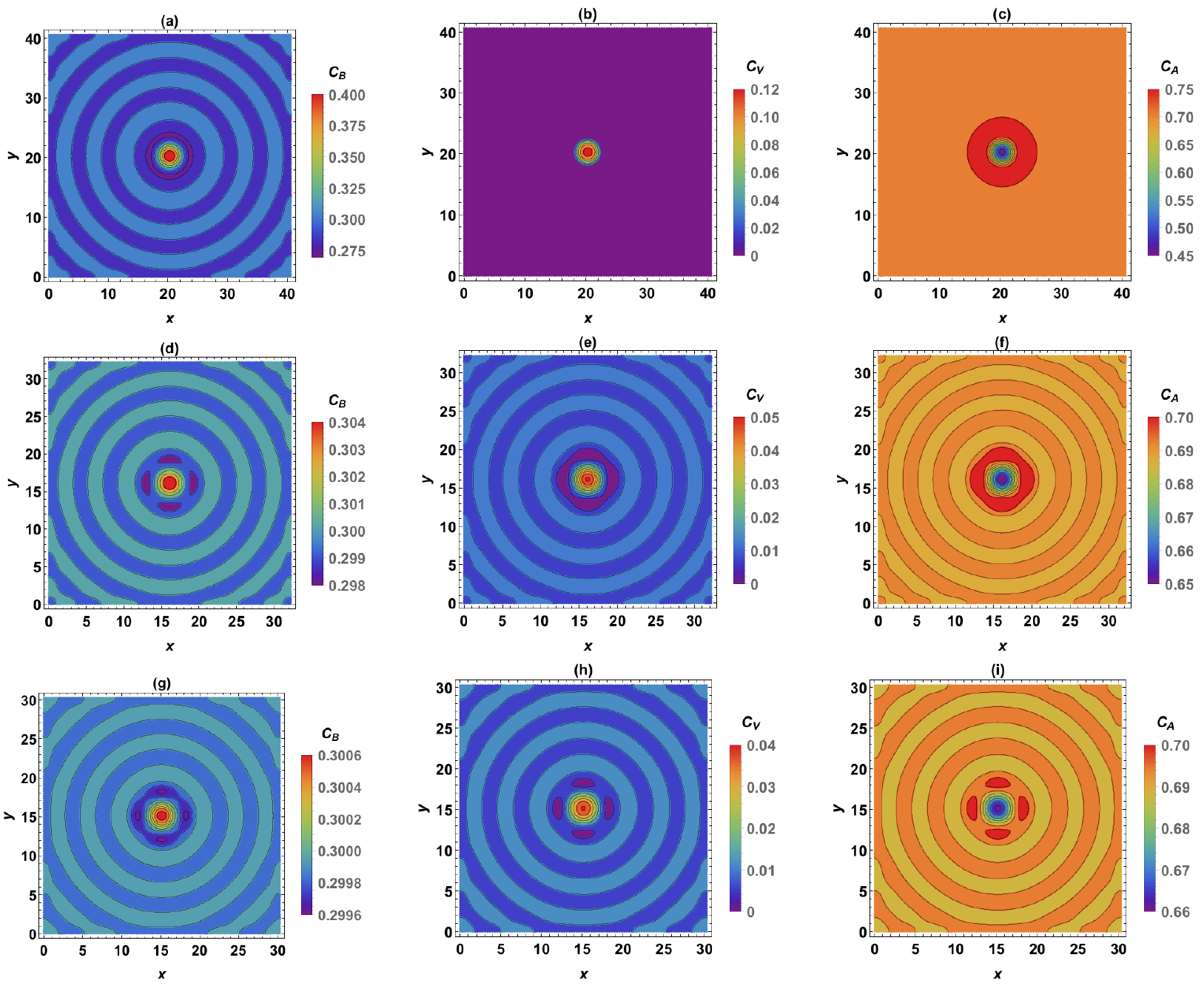}
\vspace{-0.15cm}
\caption{(a), (b), (c):  Final $C_B,\ C_V$ and $C_A$ at $\Gamma=1$. (d), (e), (f):  Final $C_B,\ C_V$ and $C_A$ at $\Gamma=10$. (g), (h), (i):  Final $C_B,\ C_V$ and $C_A$ at $\Gamma=50$. 
$Q=0.005$. 
}
\label{Fig3}
\end{figure}

Figure \ref{Fig3} shows the final concentration fields at $\Gamma=1,\ 10$ and 50 just above the threshold of a phase separation instability ($Q=0.005$). 
A close inspection reveals very different outcomes. A common feature is the formation of a vacancy denuded zone around the vacancy island, which had shrunken in size. 
At $\Gamma=1$ this zone extends to the entire exterior of the island (Fig. \ref{Fig3}(b)). Vacancies flow to the center of the island from the 
exterior; at $\Gamma=1$ the vacancy concentration increases ten-fold at the island center. This is accompanied by the $C_B$ increase at the island center from
0.3 to 0.4 (33\% increase) and a similar sized $C_A$ decrease from 0.685 to 0.45 (Fig. \ref{Fig3}(a,c)). In contrast, on the periphery of the vacancy island 
$C_B$ decreases and $C_A$ increases. The phase separation is evident in Figures \ref{Fig3}(a,c).

As $\Gamma$ increases, the vacancy denuded 
zone forms a ring around the island (Fig. \ref{Fig3}(e)), and the ring breaks into several sections with the further $\Gamma$ increase  (Fig. \ref{Fig3}(h)). 
Simultaneously, the vacancy concentration at the island center increases more moderately (but still by the factor of four in Fig. \ref{Fig3}(h)), and
$C_B$ deviates less from the initial value 0.3 as $\Gamma$ increases. In Fig. \ref{Fig3}(g) it equals to the initial value everywhere.
$C_A$ still experiences a small decrease from 0.685 to 0.66 at the island center (3.6\% difference, Fig. \ref{Fig3}(i)) and the increase to 0.7 
in the ring sections where the vacancy concentration dropped to zero. Overall, at $\Gamma=50$ the phase separation is effectively terminated.         

\begin{figure}[H]
\vspace{-0.2cm}
\centering
\includegraphics[width=6.5in]{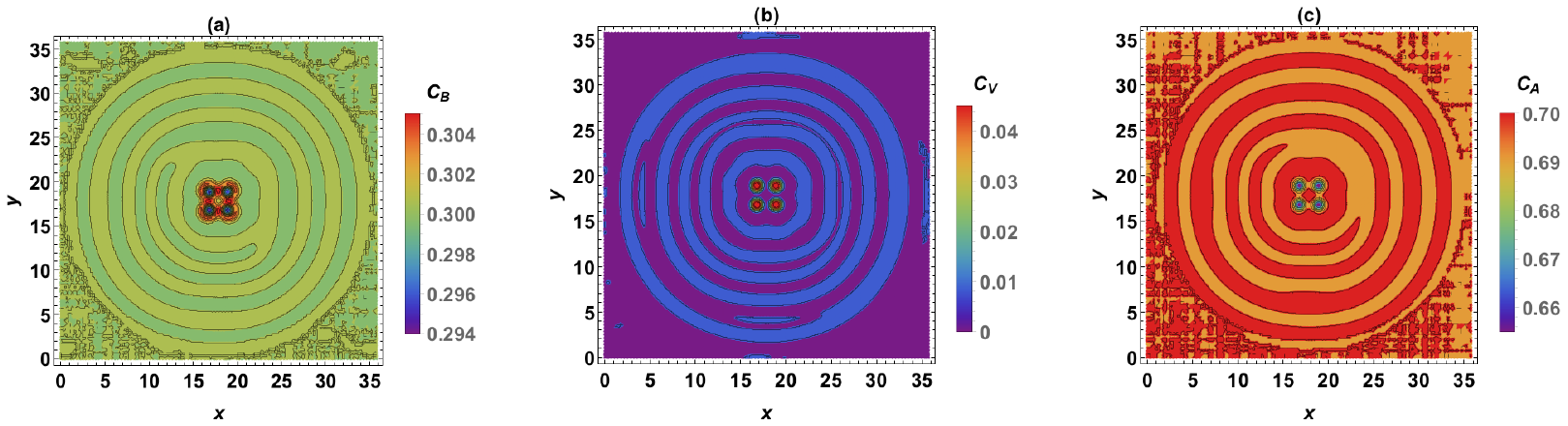}
\vspace{-0.15cm}
\caption{(a), (b), (c):  Final $C_B,\ C_V$ and $C_A$ at $\Gamma=0.002$. 
$Q=0.005$. 
}
\label{Fig4}
\end{figure}

Decreasing $\Gamma$ does not have the same effect as increasing it. In Fig. \ref{Fig4}(a,c) it is seen that at $\Gamma=0.002$ the phase separation is comparable 
to Fig. \ref{Fig3}(d,f). However, in the latter figure $\Gamma$ was increased only ten-fold from the neutral value $\Gamma=1$, whereas in the former 
figure $\Gamma$ was decreased from that value by the factor of five hundred.   
Also there appear the small-scale features in the $C_B$ and $C_A$ concentration fields inside the boundary of the vacancy island. The vacancy island itself has 
broken into four smaller islands, where $C_V$ is four times larger than the initial value. 

\begin{figure}[H]
\vspace{-0.2cm}
\centering
\includegraphics[width=6.5in]{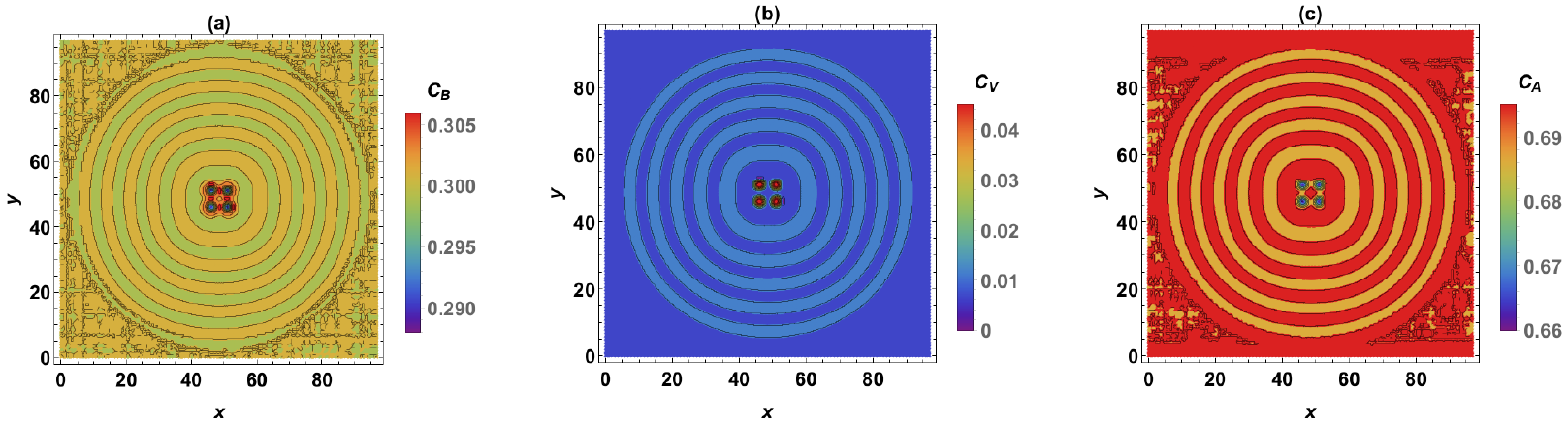}
\vspace{-0.15cm}
\caption{(a), (b), (c):  Final $C_B,\ C_V$ and $C_A$ at $\Gamma=1$. 
$Q=0.05$. 
}
\label{Fig5}
\end{figure}

Finally, one can observe in Fig. \ref{Fig5} that moving away from the instability threshold via increasing $Q$, while keeping $\Gamma$ equal to the neutral
value, has the effect similar to the one in Fig. \ref{Fig4}. Predictably, the phase separation diminishes, but it does not cease completely.

To summarize, the computations based on our chemical diffusion model of a typical 2D fcc surface alloy with a low vacancies concentration demonstrate 
that a phase separation instability is suppressed when the ratio of the jump rates $\Gamma_A/\Gamma_B$ is around fifty, 
provided that A is the more abundant atomic species. 
 The phase separation is present, 
but it is very mild (on the order of 1-2\%) when A is the slow or very slow diffusing atomic species.





\end{document}